\documentclass[aps,prd, preprint,
showpacs,showkeys,amsmath,amssymb]{revtex4}
\usepackage{xcolor}
\usepackage{graphicx}
\begin{document}

\title{Pair creation by collision of intense laser pulse with high-frequency photon beam}
\affiliation{National Research Nuclear University MEPhI, 115409 Moscow, Russia}
\author{A.~M. Fedotov} \affiliation{National Research Nuclear University MEPhI, 115409 Moscow, Russia}
\author{A.~A. Mironov} \affiliation{National Research Nuclear University MEPhI, 115409 Moscow, Russia} 
\date{October 26, 2013}

\begin{abstract}
We consider pair creation by collision of a couple of counterpropagating electromagnetic pulses with arbitrary frequency ratio, mostly in the context of setup with collision of optical and coherent hard X- (or gamma-) ray pulses. This problem is non-perturbative and in general does not admit exact analytical solutions. We discuss several known approaches according to the ratio of the parameters. Certain regions of the parameter space are not covered by the existing approximations or models. We present a new simplified exactly solvable model with one of the pulses being a delta-pulse. This model partially fills the yet unexplored gap in the space of parameters. The shape of the momentum distribution of the created particles in such a model is discussed. Even though the model may not have immediate implications for the forthcoming experiments, it can still provide some hints for better understanding of the fully non-perturbative regime and vacuum instability in QED interactions of extremely strong and short laser pulses.
\end{abstract}
\maketitle

\section{Introduction}

One of the most intriguing predictions of intense field QED is $e^+e^-$ vacuum pair production. Pair creation in a constant electric field would become observable when the field strength reaches $E_c=m^2c^3/e\hbar =1.32\cdot10^{16}\text{~V/cm}$ \cite{Sauter}. Attaining such a field strength with a constant electric field is an extremely difficult task, so it is more reasonable to use alternating, e.g. laser fields. Current progress in technologies brings hopes on attaining laser fields close to $E_c$ under the laboratory conditions. Intensities about $2\cdot 10^{22}\text{~W/cm}^2$ can be already obtained nowadays \cite{Yanovsky}, and there exist projects proposing further ways to achieve even higher intensities up to $10^{26}\text{~W/cm}^2$ \cite{ELI, XCELS}. 

Critical field $E_c$ corresponds to laser intensity $I_c\sim 10^{29}\text{ W/cm}^2$, but combining of two or more head-on colliding focused laser pulses allows to reduce the threshold intensity required for observation of pair creation down to $10^{25}-10^{26}\text{ W/cm}^2$ \cite{colliding1,colliding2}. In such schemes of experiments the laser field is often supposed to be in optical range, but it seems reasonable to take into account a situation of counterpropagating pulses with arbitrary frequencies, e.g., collision of an optical pulse with coherent hard X-ray or $\gamma$-ray pulses as proposed in \cite{Dunne}. Such coherent pulses could be obtained by different ways: by using a free-electron laser (nowadays one may think of the parameters of XFEL \cite{XFEL}), or rather by using some novel technologies of compression, e.g. the Relativistic Flying Mirror concept \cite{FlyingMirror} (see \cite{Tajima-Mourou} for a brief review).

In a most general case one may consider short high-frequency pulses of arbitrary intensity. Next it is important to choose the appropriate theoretical approach for dealing with such pulses. For the sake of simplicity we can consider the optical laser field to a high precision as an external classical field, e.g. a plane wave, due to a great number of coherent soft photons it contains. Then, to consider the problem accurately, one should consider a short high-frequency pulse interacting with that optical field as a bunch of coherent photons, so that any number (few or many) of photons from such a pulse can contribute to the process of pair formation. In other words, one should consider the problem fully quantum mechanically with respect to the short high-frequency pulse, but such a general approach as for now can not be implemented explicitly for technical reasons. Summarizing the existing methods that we list and discuss below, as for now one can consider the short high-frequency pulse either as a bunch of individual (non-coherent) photons, i.e. take into account only diagrams involving a single hard photon and any number of soft optical photons, or, if the density of hard photons is large, as another external classical field, e.g., a counterpropagating plane wave. Such two conceivable ways of consideration of short high-frequency pulse have their own benefits and limitations.

The total probability of pair production in different field configurations can be expressed in terms of the dimensionless invariant parameters. When dealing with the field configuration described above one should choose such parameters carefully. We introduce the relevant parameters and review the known approaches to the problem and their limitations in terms of the chosen parameters in Sec.~\ref{sec-params}. The approaches that we mention do not cover the whole region of the parameters of interest, so that some novel methods are needed in the currently unexplorable domains. In order to understand better matching between the two above mentioned complementary descriptions of a short high-frequency pulse in calculation of the pair production probability we present some unifying estimates based on the quasiclassical method.

In Sec.~\ref{sec-exact} we introduce a toy model which involves an idealized ultra-short pulse of electromagnetic radiation with the field strength described by a delta-function propagating opposite to the constant crossed field, the latter corresponds to the optical laser field, so that pairs are created solely in the region of an overlap. For the sake of simplicity we consider production of scalar particles, but this should not change principally the behaviour of the pair creation probability as compared to the case of fermion particles. In this paper we derive the exact analytical expression for the number of created pairs for arbitrary values of the classical non-linearity  parameter of the ultra-short pulse, and this is the most remarkable feature of the toy model. Moreover, as is shown in Sec.~\ref{sec-comparison}, the well-explored approaches mentioned above arise naturally from our exactly solvable model as special limiting cases. Finally, in Sec.~\ref{sec-conclusion} we discuss the results and indicate some prospects for possible future studies.

\section{Setup, parameters and matching of known approaches}
\label{sec-params}

Let us discuss the parameters that control the pair creation process. Assume that pairs are created  by a head-on collision of plane wave pulses. We mostly demand the case of a strong optical pulse colliding with a very short (and hence, assembled from mostly rather high-frequency Fourier components) pulse. Let $A_{L\mu}$ be a 4-potential of a strong classical external field of optical frequency. If the short pulse is described by a classical 4-potential $A_{s\mu}$, then one should analyse the process of pair creation by the resulting field $A_\mu=A_{L\mu}+A_{s\mu}$, $F_{\mu\nu}=F_{L\mu\nu}+F_{s\mu\nu}$, where $F_{\mu\nu}$ is the tensor of the electromagnetic field. Assume the pulses propagate along and against $z$-axis with $k^\mu_L=\{\omega_L;\,0,\,0,\,\omega_L\}$ and $k^\mu_s=\{\omega_s;\,0,\,0,\,-\omega_s\}$ respectively, so that $k_L\cdot A_L=k_s\cdot A_s=0$. Then the resulting field is described by the dimensionless parameters 
\begin{equation}
\label{eps-eta}
\varepsilon,\eta=\frac{e}{m^2}
\sqrt{\sqrt{(F_L\cdot F_s)^2+(F_L\cdot F_s^*)^2}\mp F_L\cdot F_s},
\end{equation}
which possess the meaning of the magnitudes of the electric $E$ and magnetic $B$ field strengths, normalized to the critical QED field\footnote{We assume $\hbar=c=1$ throughout the paper.} $E_c=m^2/e$, in such a reference frame that $\vec{E}\Vert\vec{B}$.

Alternatively, if one prefers to describe the short pulse as a bunch of individual photons with energy $\omega_s$, propagating in the external field $A_L$ ($\omega_L\ll\omega_s$), then the probability of pair creation would depend on the following set of invariant parameters \cite{Ritus}: 
\begin{equation}
\label{a-kappa}
\xi_{L,s}=\frac{e}{m}\sqrt{-A_{L,s\,\mu}^2}=\frac{eE_{L,s}}{m\omega_{L,s}},\quad
\varkappa_{L,s}=\frac{e}{m^3}\sqrt{-(F_{s,L}\cdot k_{L,s})^2}.
\end{equation}
The parameters $\xi_{L}$, $\xi_{s}$ are the classical non-linearity parameters and are proportional to the square roots of the photon densities. Parameters $\varkappa_{L,s}$ are the dynamical quantum parameters, larger values of $\varkappa$ correspond to higher energy of photons in a beam. In the most of the paper we will consider collision of low-frequency electromagnetic pulse $A_{L}$ with high-frequency $A_{s}$ so that $\omega_L\ll\omega_s$, and supposing $A_L$ is intense, we assume $\xi_{L}\rightarrow\infty$ which means that $A_L$ may be regarded as constant crossed field \cite{Ritus}. Also, under the same assumption we neglect $\varkappa_L$ assuming $\varkappa_L\rightarrow 0$. For $\varkappa_s$ in a field configuration under discussion we have: $\varkappa_s=2eE_L\omega_s/m^3$. In case of extremely short subperiod pulses, $\omega_s$ means the largest typical frequency they are assembled from, i.e. $\varkappa_s\sim 2eE_L/\tau_sm^3$, where $\tau_s$ is the period of a pulse.

Probability of pair creation in strong electromagnetic fields was obtained for different values of parameters (\ref{eps-eta}) and (\ref{a-kappa}). If $\xi_{L},\xi_{s}\gg 1$ and $\varkappa_{L,s}\ll \varepsilon,\,\eta$, then both $A_L$ and $A_s$ may be treated classically and the resulting field can be considered locally constant and homogeneous \cite{focused, colliding1, Fedotov}. This approach is fully non-perturbative and takes into account absorption of great number of photons from \textit{both} fields during the process of pair creation. In this case probability is determined mainly by the field strength and not by the field frequencies, so that the case of high frequencies cannot be treated in an appropriate way. This approximation is limited by a value of field inhomogeneity, so that $l_i\gg l_C$ where $l_i$ is a characteristic size of variation of the field and $l_C$ is the electron Compton length. $l_i$ is related to the frequency of a field, so that frequencies $\omega\gtrsim m$ are prohibited within such an approach.

Assuming the short pulse to be a bunch of high-frequency non-coherent photons, in order to find the number of created pairs one should calculate the probability of pair photoproduction in a strong external field. The case of pair photoproduction by photons of arbitrary frequencies in the plane wave $A_L$ was well studied in \cite{photon-in-plane-wave}. That approach was perturbative with respect to high-frequency pulse, since only diagrams with a single hard photon line were taken into account for the process. To obtain a total pair production rate, the probability of pair photoproduction by a single photon must be multiplied by the density $n_\gamma$ of hard photons, and after that integrated over the region of overlapping of the beams. The parameters $\xi_{L}$ and $\varkappa_s$ in this case may be arbitrary, but due to perturbative (in the abovementioned sense) treatment $\xi_{s}\ll 1$. If $\xi_{L}\rightarrow\infty$, then $A_L$ becomes a constant crossed field \cite{Ritus}, such a case is rather convenient to study the dependency of the probability on $\varkappa_s$:
\begin{equation}
W_{e^-e^+}\sim\left\{\begin{array}{lr}\displaystyle
0.23\frac{e^2m^2}{\omega_s}  \varkappa_s e^{-8/3\varkappa_s},&\varkappa_s\ll 1;\\\\
0.38\frac{e^2m^2}{\omega_s}\varkappa_s^{2/3},&\varkappa_s\gg 1.\end{array}\right.
\end{equation}
Pair creation by a photon in a constant electromagnetic field is also well studied using the expressions for imaginary part of polarization operator in constant and homogeneous electric field \cite{Narozhny-polarization, Baier, Dunne}, as well as in constant and homogeneous electromagnetic field \cite{Katkov}. However, such an approach can hardly be repeated for a process involving larger number of photons from $A_s$, partially due to rapid increasing of complexity of the resulting expressions (see e.g.~\cite{BKS} for details), and partially due to a less direct relation between the loop diagrams with higher number of external legs and the total pair creation probability via the optical theorem.

There is yet another approach to the problem originally developed in \cite{Brezin, Popov} and based on a quasiclassical approximation. It can handle both perturbative and non-perturbative regimes, but is valid still only for the case of low frequencies $\omega_{L,s}\ll m$.

\begin{figure*}
\includegraphics[width=4in]{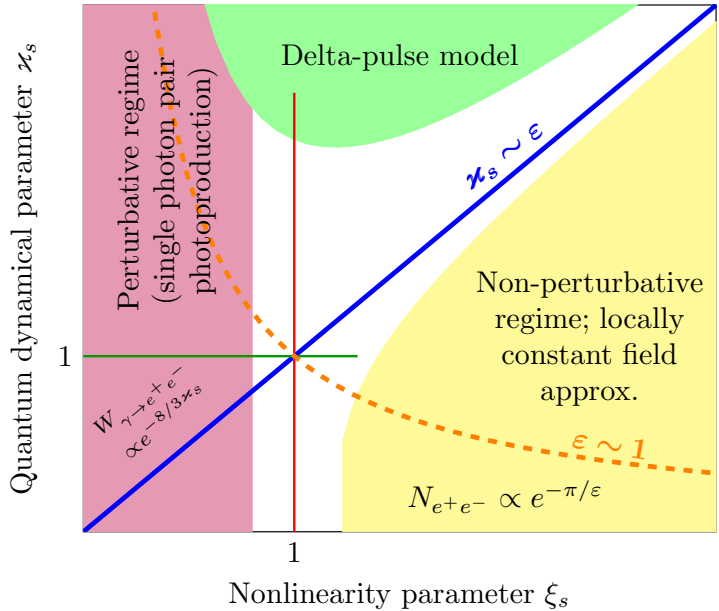}
\caption{Layout of theoretical models in the plane of parameters $(\xi_{s},\varkappa_s)$. Coloured regions correspond to the domains of applicability of different approaches (each marked by its caption), dashed line indicates the critical field limit $E_L\sim E_c$.}
\label{landscape}
\end{figure*}

Obviously, in order to put different approaches together and discuss them on unified grounds, an important issue would be establishing the complete set of relevant parameters for the problem. Let us choose $\xi_{s}$ and $\varkappa_s$ as the main parameters. For the sake of simplicity, assume a particular case of polarization of the pulses $A_L$ and $A_s$, such that the electric and magnetic fields of the two waves interfere constructively and destructively, respectively. Then the parameters (\ref{eps-eta}) can be expressed in terms of (\ref{a-kappa}), namely 
\begin{equation}\label{eps-eta1}
\varepsilon=\frac{e}{m^2}\sqrt{E^2-B^2}
=\frac{e}{m^2}\sqrt{\left(E_L+E_s\right)^2-\left(E_L-E_s\right)^2}=\frac{2e(E_LE_s)^{1/2}}{m^2}=\sqrt{2\xi_{s}\varkappa_s},\quad \eta=0.
\end{equation}
Now it becomes possible to dispose the domains of applicability of each of the approaches at the same diagram in the plane $(\xi_{s},\varkappa_s)$ (see Fig. \ref{landscape}). The approximation of locally constant and homogeneous field is applicable only for $\xi_{s}\gg 1$ and $\varkappa_s \ll \varepsilon$ (yellow region on Fig. \ref{landscape}), while 
perturbative approach is applicable for arbitrary $\varkappa_s$, but only if $\xi_{s}\ll 1$. One can see that the strip $\xi_{s}\sim 1$, which for photon energies in the range of hard X-rays or $\gamma$-rays 
corresponds to a rather intense field, is not covered by any of the existing approaches, so that there is no model connecting together these approaches. Also, the region of $\xi_{s}\gg 1$ and $\varkappa_s \gg \varepsilon\gg 1$ is not described appropriately --
it is about the super-critical fields $E\gg E_c$, which are seemingly non-physical \cite{laser-limit}, and generally anyway requires taking the backreaction of the created electron-positron plasma into account.

In this paper we do not intend to solve the whole general problem posed above. In the rest of the section, let us present a rather simple approach for $\xi_s\ll 1$ and $\varepsilon\ll 1$, which allows to match the known asymptotic expressions of the generally rather complicated expression in the perturbative region (to cross the blue sloping line $\varkappa_s\sim\varepsilon$ in the left rectangular colored region on Fig. \ref{landscape}). For this, we use a version of the quasiclassical method.  The probability $W_{e^-e^+}$ of pair creation is proportional to 
\begin{equation}
\label{w-prob}
W_{e^-e^+}\propto\exp\left\{-2\,{\rm Im}\int_0^{t_*}[{\cal E}_f(t)-{\cal E}_i(t)]\,dt\right\},
\end{equation}
where integration is over the moment of occurrence of the process, ${\cal E}_{i,f}$ are the initial and final energies of the system, ${\cal E}_f(t)-{\cal E}_i(t)$ is the energy release in the process and the upper limit $t_*$ is defined by the stationary phase condition ${\cal E}_f(t_*)={\cal E}_i(t_*)$.  

Consider first a photon with energy and momentum $\omega_s$, propagating transversely in a constant electric field $E_L\ll E_c$. Parameters $\varepsilon$ and $\varkappa_s$ for such a case are $\varepsilon=eE_L/m^2$, $\varkappa_s=eE_L\omega_s/m^3$. We have only a single photon in the initial state, ${\cal E}_i=\omega_s$, while in the final state we assume a pair to be created instead of a photon, so that there are an electron $e^-$ and a positron $e^+$, for simplicity assume that they possess equal \footnote{It is not hard to see that such outcome indeed corresponds to the extremum of the expression in exponential.} momenta $\omega_s/2$. They subsequently start being accelerated by the electric field, so that ${\cal E}_f(t)=2\sqrt{m^2+(\omega_s/2)^2+(eE_Lt)^2}$. In this case $t_*=im/eE_L$. By integrating in (\ref{w-prob}) we obtain
\begin{widetext}
\begin{equation}
\label{w-electric-L}
W_{e^-e^+}\propto\exp\left\{-\frac{2m^2}{eE_L}\left[\left(1+\left(\frac{\omega_s}{2m}\right)^2\right)\,{\rm arctan}\left(\frac{2m}{\omega_s}\right)-\frac{\omega_s}{2m}\right]\right\},
\end{equation}
or, expressing the result in terms of the invariant parameters $\varepsilon$ and $\varkappa_s$,
\begin{equation}
\label{w-electric}
W_{e^-e^+}\propto\exp\left\{-\frac{2}{\varepsilon}\left[\left(1+\left(\frac{\varkappa_s}{2\varepsilon}\right)^2\right)\,{\rm arctan}\left(\frac{2\varepsilon}{\varkappa_s}\right)-\frac{\varkappa_s}{2\varepsilon}\right]\right\}\sim\left\{\begin{array}{lr} e^{-\pi/\varepsilon},& \varkappa_s\ll\varepsilon\ll 1;\\ e^{-8/3\varkappa_s},&\varepsilon\ll\varkappa_s\ll 1.\end{array}\right.
\end{equation}
\end{widetext}
This estimate demonstrates an expected behaviour of $W_{e^-e^+}$ in both regimes of high and low $\varkappa_s$: if the energy of the initial photon is low ($\varkappa_s\ll\varepsilon\ll 1$), then pair production from vacuum by the electric field $E_L$ totally dominates, and one can think that the initial photon just gets absorbed by one of created particles. Alternatively, if the photon energy increases so that $\varepsilon\ll\varkappa_s\ll 1$, then the pair creation process becomes indeed induced by the initial photon. More sophisticated derivations of this result based on the explicit expression for polarisation operator in a plane wave field \cite{Narozhny-polarization} can be found in \cite{Katkov,Baier,Dunne}.

After we have carried out the estimation (\ref{w-electric}), let us come back to a more complicated case of our primary interest, i.e. to two counterpropagating pulses. Our method here may resemble in a sense a mean-field approach. Assume that the pulses are polarized as above and that both of them may be described locally as constant crossed EM field ($\xi_{L},\xi_{s}\gg 1$), so that the resulting field is $\vec{E}=\left\{E_L+E_s,0,0\right\}$, $\vec{B}=\left\{0,E_L-E_s,0\right\}$ (we assume $E_L>E_s$) and the parameters (\ref{eps-eta}) are as in (\ref{eps-eta1}). Let us pick out and consider a particular single photon with energy $\omega_s$ participating the field $E_s$. It propagates transversely in the resulting field and is characterised by $\varkappa_s$. One can mentally split the pair creation process as been invoked by two mechanisms: either by the resulting constant electromagnetic field from vacuum, or as induced by the photon under consideration (if its energy is high enough) in the same field. All the parameters $\varepsilon,\;\varkappa_s, \xi_{L,s}$ are invariant and we can change the reference frame so that the magnetic field becomes zero, $\vec{B}^\prime=0$. To achieve this, the velocity of a new reference frame should be directed along $z$-axis and its magnitude must be $V=\frac{E_L-E_s}{E_L+E_s}$. The electric field in the new reference frame is $E^\prime=2\sqrt{E_LE_s}$ and the photon frequency is Doppler-shifted as $\omega^\prime_s=\sqrt{E_L/E_s}\,\omega_s$. In this new reference frame the estimation derived above (\ref{w-electric-L}) can be applied, and after substitution of $E^\prime$ and $\omega^\prime_s$ we obtain
\begin{equation}
\label{w-em}
W_{e^-e^+}\propto\exp\left\{-\frac{2}{\varepsilon}\left[\left(1+\left(\frac{1}{n}\right)^2\right)\,{\rm arctan}\left(n\right)-\frac{1}{n}\right]\right\},
\end{equation}
where $n$ denotes 
\begin{equation}
n=\frac{2m}{\omega^\prime_s}=\frac{2m}{\omega_s}\sqrt{\frac{E_s}{E_L}}=2\sqrt{\frac{2\xi_{s}}{\varkappa_s}},
\end{equation}
i.e. may be interpreted as the number of photons that must be absorbed from the field $A_s$ in order to create a pair (this is an invariant quantity). As in (\ref{w-electric}), in the limiting cases we have $W_{e^-e^+}\propto e^{-\pi/\varepsilon}$ for $n\gg 1$ and $W_{e^-e^+}\propto e^{-8/3\varkappa_s}$ for $n\ll 1$. Thus the denominator in the exponential is defined by the largest of the two parameters $\varepsilon$ and $\varkappa_s$. This is actually the reason why harder pulses (with $\varkappa_s\gtrsim\varepsilon$) can stimulate pair production.

One can see that the actual asymptotic expression actually depends on the ratio of non-linearity parameter $\xi_{s}$ and the quantum dynamical parameter $\varkappa_s$. In particular, the locally constant field approximation is valid as long as $\varkappa_s\ll\xi_s$ (besides the other requirements formulated above). Of course, our oversimplified  consideration is not capable for calculation of a preexponential factor, which must also be a function of the invariant parameters. 

\section{Exact solution for collision with a delta-pulse}
\label{sec-exact}

As was shown in the preceding section, there exists a region on the plane of parameters $(\xi_{s},\,\varkappa_2)$ for the problem of pair creation by a collision of electromagnetic pulses, for which 
none of the yet proposed approaches capable of explicit quantitative calculation of the number of created pairs can be applied. On the one hand, to the best of our knowledge exact solutions for such a setup have been lacking. On the other hand, the imaginary time method is too restrictive and qualitative, because does not take into account spatial variation of the field. Hence, let us introduce an exactly solvable model, which is valid for arbitrary values of $\xi_s$. Namely, consider a delta-pulse of electromagnetic radiation (a "hard" pulse) counterpropagating to an arbitrary plane wave (a "soft" pulse). For simplicity, in this section we consider creation of scalar pairs, but generalization to creation of electron-positron pairs seems to be an easy task.

Consider a massive charged scalar field in an external electromagnetic background. The field obeys the Klein-Gordon equation:
\begin{equation}
\label{KGequation}
\{[\partial_\mu-ieA_\mu(x)]^2+m^2\}\Psi(x)=0.
\end{equation}
If $A_\mu(x)$ is a plane wave propagating along $z$-axis, i.e. $A_{\mu}(x)=A_{\mu}(\varphi)$, where $\varphi=kx=\omega x_{-}$, $x_{-}=t-z$ and $k\cdot A=0$, then (\ref{KGequation}) can be solved in terms of Volkov solutions \cite{Volkov,Zakowicz,Madalina}
\begin{equation}
\label{Volkov}
\Psi_{\vec{p}_\perp,p_-}=\frac1{\sqrt{2|p_-|(2\pi)^3}}\exp\left\{-ip x+\frac{ie}{k p}
\int_0^{k\cdot x}\left[p A(\varphi)+\frac{e}{2}A^2(\varphi)\right]\,d\varphi\right\},
\end{equation}
where $p^\mu$ ($p^2=m^2$) is 4-momentum of a scalar particle. The quantum number $p_-=p^0-p^3$ is a quantity conserving in the plane wave, $\vec{p}_\perp$ is a projection of the momentum onto the $xy$-plane and $p^0=\frac12\left(p_-+\frac{m^2+p_\perp^2}{p_-}\right)$ is the energy of the particle. It is well known that the solutions (\ref{Volkov}) are complete, normalized and satisfy the conditions of orthogonality (see the Appendix and \cite{Madalina} for details):
\begin{equation}
\label{normalization}
\int d^2x_\perp dx_- \Psi_{\vec{p}_\perp,p_-}^*\left(i\stackrel{\leftrightarrow}{\frac{\partial}{\partial x_-}}-eA_+\right)\Psi_{\vec{p}_\perp^{\,\prime},p_-'}
={\rm sgn}(p_-)\delta(\vec{p}_\perp-\vec{p}_\perp^{\,\prime})\delta(p_--p_-'),
\end{equation}
where $f\stackrel{\leftrightarrow}{\frac{\partial}{\partial x_-}}g\equiv f\frac{\partial}{\partial x_-}g-g\frac{\partial}{\partial x_-}f$. The values $p_->0$ and $p_-<0$ correspond to positive and negative energies, respectively. The second quantized field $\Psi(x)$ can be expanded as
\begin{equation}
\label{field_expansion}
\Psi(x)=\int\limits d^2p_\perp \int\limits_0^\infty dp_-\left\{\Psi_{\vec{p}_\perp,p_-}(x)\,a_{\vec{p}_\perp,p_-}+\Psi_{-\vec{p}_\perp,-p_-}(x)\,b_{\vec{p}_\perp,p_-}^\dagger\right\},
\end{equation}
where $a_{\vec{p}_\perp,p_-}$ and $b_{\vec{p}_\perp,p_-}$ are the annihilation operators for scalar particles and antiparticles, respectively.

A counterpropagating delta-pulse of electromagnetic field may be described by 4-potential ${\cal A}_\mu={\cal A}_{0\mu}\theta(x_+)$, where $x_+=t+z$ and the step function $\theta(x_+)=1$ for $x_+>0$ and $\theta(x_+)=0$ for $x_+<0$. Then the total resulting 4-potential is
\begin{equation}
A_\mu(x)=A_{L\mu}(x_-)+{\cal A}_{0\mu}\,\theta(x_+),
\end{equation}
where by $A_{L\mu}(x_-)$ we denote the original plane wave. The world line $x_+=0$ separates the spacetime into the two regions, and inside each of them the electromagnetic field is just a plane wave alone. 
Since the plane wave alone is not capable for pair creation, the pairs are created only at the boundary $x_+=0$ between these regions. The region $x_+<0$ (before the delta-pulse has arrived) can be identified with the "in-region", similarly $x_+>0$ (after the delta-pulse has passed) is the "out-region". With respect to such interpretation let us introduce the in- and out-modes $\Psi^{(in)}_{\vec{p}_\perp,\,p_-}(x)=\Psi_{\vec{p}_\perp,\,p_-}(x;A_L)$, $\Psi^{(out)}_{\vec{p}_\perp,\,p_-}(x)=\Psi_{\vec{p}_\perp,\,p_-}(x;A_L+{\cal A}_0)$ ($p_->0$). Having the sets of in- and out-modes one can write the expansion (\ref{field_expansion}) in each of the regions, thus introducing the in- and out- creation and annihilation operators.

In our model, the in- and out- modes are related by the matching condition at $x_+=0$:
\begin{equation}\label{matching}
\Psi_{\vec{p}_\perp^{\,\prime},\,p_-'}(x;A_L)\vert_{x_+=0}=\int d^2p_\perp\int_{-\infty}^{+\infty} dp_-\,\alpha_{\vec{p}_\perp,\,p_-; \vec{p}_\perp^{\,\prime},\,p_-'}\,\Psi_{\vec{p}_\perp,\,p_-}(x;A_L+{\cal A}_0)\vert_{x_+=0}.
\end{equation}
Accordingly, one obtains the Bogolubov transformation relating the in- and out-operators:
\begin{equation}\label{Bogolubov}
a_{\vec{p}_\perp,p_-}^{(out)}=\int\limits d^2p_\perp\int\limits_0^\infty dp_-'\,\left\{\alpha_{\vec{p}_\perp,\,p_-; \vec{p}_\perp^{\,\prime},\,p_-'} a_{\vec{p}_\perp',p_-'}^{(in)}   +\alpha_{\vec{p}_\perp,\,p_-; -\vec{p}_\perp^{\,\prime},\,-p_-'} b_{\vec{p}_\perp',p_-'}^{(in)\dagger}\right\}.
\end{equation}
The coefficients $\alpha_{\vec{p}_\perp,\,p_-; \vec{p}_\perp^{\,\prime},\,p_-'}$ are thus the Bogolubov coefficients; once they are found the total amount of pairs created in the process takes the form
\begin{equation}
\label{N}
N_{e^+e^-}=\int\limits d^2p_{\perp}\,\int\limits_0^\infty dp_- \langle 0_{in}|a_{\vec{p}_\perp,p_-}^{(out)\dagger}a_{\vec{p}_\perp,p_-}^{(out)}|0_{in}\rangle =
\int\limits d^2p_{\perp}\int\limits_0^\infty dp_-\int\limits d^2p_{\perp}'\int\limits_{-\infty}^{0} dp_-' |\alpha_{\vec{p}_\perp,\,p_-; \vec{p}_\perp^{\,\prime},\,p_-'}|^2.
\end{equation}

Before proceeding further with calculation, let us elaborate a bit more on the particular form of the electromagnetic fields. Since both fields are propagating parallel to $z$-axis, it is convenient to choose the gauge $A_0=A_3=0$. For the sake of simplicity we assume in what follows that the delta-pulse is linearly polarized along the $x$-axes, $\vec{{\cal A}}_0=\left\{-{\cal A}_0,0,0\right\}$, and that $A_L$ is a constant crossed field (this is locally a rather good approximation if this is an optical laser field, see sec. \ref{sec-params}). As previously, we assume that relative polarization of the fields is such that their electric fields are summed up, whereas the magnetic fields are subtracted:  $\vec{A}_L=\left\{-E_Lx_-,0,0\right\}$. Then the only non-zero components of the resulting EM field are 
\begin{equation}
\label{result-field}
E_x=E_L+{\cal A}_0\delta(x_+),\quad B_y=E_L-{\cal A}_0\delta(x_+).
\end{equation}
The field strength of a delta-pulse is infinite, $E_\gamma\vert_{x_+=0}={\cal A}_0\delta(0)=\infty$,
and because of this when necessary we need to introduce a regularization. Assuming that the actual duration of the "hard" pulse $\tau$ is the smallest time parameter in the problem, we can treat the expression for its field strength as 
\begin{equation}
\label{physical-meaning}
E_\gamma={\cal A}_0/\tau.
\end{equation}
The field invariants (\ref{eps-eta}) for our field (\ref{result-field}) are
\begin{equation} 
E^2-B^2=4E_L{\cal A}_0\delta(x_+)\ge 0,\quad
\vec{E}\cdot\vec{B}=0,\quad \varepsilon\vert_{x_+=0}\sim \frac{2e\sqrt{E_L E_\gamma}}{m^2},\quad \eta=0.
\end{equation}
In the expanded form, the Volkov solutions with our field configuration (\ref{result-field}) can be written as
\begin{equation}
\begin{split}
\label{explicit-Volkov}
\Psi_{\vec{p}_\perp,\,p_-}=\frac{1}{\sqrt{2|p_-|(2\pi)^3}}\,\exp & \left\{ i\vec{p}_\perp\cdot\vec{x}_\perp-i\frac{(m^2+p_\perp^2)x_-}{2p_-}-i\frac{p_-x_+}{2}\right.\\
&-\left. i\frac{e}{p_-}\int_0^{x_-}\left[\vec{p}_\perp\cdot\vec{A}_L(x_-)+\frac{e}2 \vec{A_L}^2(x_-)\right]\,dx_-\right\}.
\end{split}
\end{equation}

With all these refinements, let us come back to evaluation of $\alpha_{p_-,\,\vec{p}_\perp; p_-'\vec{p}_\perp^{\,\prime}}$. Using (\ref{matching}) and the orthogonality and normalization condition (\ref{normalization}), it is possible\footnote{In what follows, we ignore all the phase factors arising in $\alpha_{p_-,\vec{p}_\perp ;p_-',\vec{p}_\perp^{\,\prime}}$, because they all the same do not contribute to the number of created pairs.} to express  $\alpha_{p_-,\vec{p}_\perp ;p_-',\vec{p}_\perp^{\,\prime}}$:
\begin{equation}
\label{alpha-int}
\alpha_{\vec{p}_\perp,\,p_-; \vec{p}_\perp^{\,\prime},\,p_-'}=i\int d^2x_\perp\int_{-\infty}^{+\infty}dx_-\left.\left[\Psi_{\vec{p}_\perp,p_-}^*(x;A_L+{\cal A}_0)\stackrel{\leftrightarrow}{\frac{\partial}{\partial x_-}}\Psi_{\vec{p}_\perp^{\,\prime},p_-'}(x;A_L)\right]\right\vert_{x_+=0}.
\end{equation} 
Let us substitute $\Psi_{\vec{p}_\perp^{\,\prime},p_-'}$ in the form (\ref{explicit-Volkov}) and take into account that we need the coefficients with $p_->0$, $p_-'<0$. For the latter, it is convenient to introduce $q_-=-p_-'>0$. Integration over $\vec{x}_\perp$ may be performed immediately and results in $(2\pi)^2\delta(\vec{p}_\perp-\vec{p}_\perp^{\,\prime})$, so that we obtain
\begin{equation}\label{alpha-int1}
\alpha_{\vec{p}_\perp,\,p_-; \vec{p}_\perp^{\,\prime},\,p_-'}=\frac{\delta(\vec{p}_\perp-\vec{p}_\perp^{\,\prime})}{8\pi\sqrt{p_-q_-}}\int_{-\infty}^{+\infty}dx_-\,P(x_-)\,\exp\left[i\Phi(x_-)\right],
\end{equation} 
where
\begin{equation}
\begin{split}
P(x_-)=&(p_\perp^2+m^2)\frac{q_--p_-}{p_-q_-}-2\frac{p_x}{p_-}{e\cal A}_0+\frac{1}{p_-}e^2{\cal A}_0^2\\
&-2\frac{q_--p_-}{p_-q_-}p_xeE_Lx_-+2\frac{1}{p_-}eE_Le{\cal A}_0x_-+\frac{q_--p_-}{p_-q_-}e^2E_L^2x_-^2
\end{split}
\end{equation}
and
\begin{equation}
\begin{split}
\Phi(x_-)=&\frac{1}{2}\left[(p_\perp^2+m^2)\frac{p_-+q_-}{p_-q_-}-2\frac{p_x}{p_-}e{\cal A}_0+\frac{1}{p_-}e^2{\cal A}_0^2\right]x_-\\
&-\frac{eE_L}{2}\left(p_x\frac{p_-+q_-}{p_-q_-}-\frac{e{\cal A}_0}{p_-}\right)x_-^2+\frac{e^2E_L^2(p_-+q_-)}{6p_-q_-}x_-^3.
\end{split}
\end{equation} 
The function $\Phi(x_-)$ is a cubic polynomial in $x_-$, so the integral in (\ref{alpha-int1}) by a substitution $x_-\to \zeta$,
\begin{equation}\label{zeta}
x_-=-\zeta\sqrt[3]{\frac{2p_-q_-}{e^2E_L^2(p_-+q_-)}}+\frac{p_x}{eE_L}-\frac{q_-{\cal A}_0}{(p_-+q_-)E_L},
\end{equation}
can be evaluated in terms of the Airy function ${\rm Ai}(w)=\frac{1}{2\pi}\int_{-\infty}^{+\infty}d\zeta\,\exp\left[-i\left(\frac{\zeta^3}{3}+w\zeta\right)\right]$ and its derivative ${\rm Ai}'(w)$, where
\begin{equation}
\label{w}
w=\frac{(p_-+q_-)^2(m^2+p_y^2)+e^2{\cal A}_0^2p_-q_-}{(2eE_Lp_-q_-)^{2/3}(p_-+q_-)^{4/3}}.
\end{equation}
For further reference, let us rewrite Eq.~(\ref{zeta}) in a different form by expressing $p_x$:
\begin{equation}\label{px}
p_x=\zeta\sqrt[3]{\frac{2eE_Lp_-q_-}{p_-+q_-}}+
\frac{q_-}{p_-+q_-}e{\cal A}_0+eE_Lx_-.
\end{equation}
Since the integral defining the Airy function is contributed mostly from $|\zeta|\lesssim 1$, it is very likely that the three successive terms on the RHS of Eq.~(\ref{px}) can be ascribed the meaning of the (random) $x$-component of momentum just after creation of a particle, an initial jerk of a particle by the field of a delta-pulse, and variation of its momentum in the course of further travelling inside the field $A_L$, respectively. To convince of such interpretation, note first that motion in the field $A_L$ alone after turning away from the delta-pulse is quasiclassical as  described by the Volkov solutions, which are of the form $e^{iS}$. The $x$- component of the classical equation of motion reads $\dot p_x=eE_x-eB_yv_z=eE_L(1-v_z)=eE_L\dot x_-$. Thus $dp_x=eE_Ldx_-$, so that $p_x=eE_L x_-+{\rm const}$, in agreement with our interpretation of the last term in Eq.~(\ref{px}).

In order to calculate the total number of created pairs, it remains to integrate $|\alpha_{p_-,\vec{p}_\perp ;p_-',\vec{p}_\perp^{\,\prime}}|^2$ over $\vec{p}$ with $p_->0$ and over $\vec{p'}$ with $p_-'<0$, as prescribed by Eq.~(\ref{N}). However, the magnitude of $\alpha_{p_-,\vec{p}_\perp ;p_-',\vec{p}_\perp^{\,\prime}}$ does not depend neither on $p_x$ nor on $p_x'$, so that formally integrals over $p_x$ and $p_x'$ are divergent. Recall that a similar feature reveals in pair creation by a constant electric field. As in the latter example, the probability distribution over $p_x$ is uniform, but the particles with given $p_x$ are all created exclusively at the location with certain $x_-$ and are all coming from there. Relating $x_-$ to the instance of creation (which obviously occures inside a delta-pulse following the world line $x_+=0$), we can write $\int dp_x=eE_L \Delta x_-\vert_{x_+=0}\sim 2eE_L\times T$, where $T$ is the total time of observation. Hence, in (\ref{N}) we have
\[\int d^2p_\perp d^2p_\perp'\delta^2(\vec{p_\perp}-\vec{p}_\perp^{\,\prime})\ldots=\int d^2p_\perp\delta(\vec{0})\ldots=\int\frac{dp_x dp_y S_\perp}{(2\pi)^2}\ldots=\int\frac{dp_y 2eE_LT S_\perp}{(2\pi)^2}\ldots\]
By taking a quotient of (\ref{N}) over $S_\perp$
and over the observation time $T$, one passes to the particle production rate per unit area of the wave front, which takes the form
\begin{widetext}
\begin{equation}
\label{numpairs1}
\begin{split}
\frac{N_{e^+e^-}}{S_\perp\cdot T}=\frac{eE_L}{2\pi^2}\int_{-\infty}^{+\infty} dp_y\int_0^\infty dp_-\int_0^\infty dq_- \frac{1}{4p_-q_-}
&\left\{\left[\frac{(2p_-q_-)^{1/3}(p_--q_-)e^2{\cal A}_0^2}{(eE_L)^{2/3}(p_-+q_-)^{7/3}}\right]^2\,{\rm Ai}^2\,(w)\right.\\
&\quad+\left.4\left[\frac{(2p_-q_-)^{2/3}e{\cal A}_0}{(eE_L)^{1/3}(p_-+q_-)^{5/3}}\right]^2{\rm Ai}'^2\,(w)\right\}.
\end{split}
\end{equation}
\end{widetext}
At this point it is convenient to change the variables $\{p_-,q_-\}\to \{u,\lambda\}$, so that $p_-=(1-\lambda)u^{-3/2}$, $q_-=\lambda u^{-3/2}$, $0<u<\infty$, $0<\lambda<1$ and the Jacobian is $|\frac{\partial(p_-,q_-)}{\partial(\lambda,u)}|=\frac32 u^{-4}$. In terms of the new variables, the expression Eq.~(\ref{w}) for $w$ becomes
\begin{equation}
w=k(\lambda)u\equiv\frac{m^2+p_y^2+e^2{\cal A}_0^2\lambda(1-\lambda)}{[2eE_L\lambda(1-\lambda)]^{2/3}}\,u.
\end{equation}
Integration over $u$ can be now performed using the formulas
\begin{equation}
\int_0^{+\infty}u\,{\rm Ai}^2(ku)\,du=\frac{1}{6\pi\sqrt{3}k^2},\quad \int_0^{+\infty}\,{\rm Ai}'^2(ku)\,du=\frac{1}{3\pi\sqrt{3}k}.
\end{equation}
Hence, the expression (\ref{numpairs1}) transforms into
\begin{equation}
\label{numpairs2}
\begin{split}
\frac{N_{e^+e^-}}{S_\perp\cdot T}=\frac{eE_L\cdot e^2{\cal A}_0^2}{8\pi^3\sqrt{3}}
&\int_{-\infty}^{+\infty} dp_y\int_0^1 d\lambda \;\lambda(1-\lambda)\\
&\times \left\{\frac{8}{[m^2+p_y^2+e^2{\cal A}_0^2\lambda(1-\lambda)]}+\frac{e^2{\cal A}_0^2(1-2\lambda)^2}{[m^2+p_y^2+e^2{\cal A}_0^2\lambda(1-\lambda)]^2}\right\}.
\end{split}
\end{equation}
We can compute the integral over $p_y$ by enclosing the integration contour in the upper half of the complex plane of $p_y$ and using the Cauchy's residue theorem, so that
\begin{equation}
\frac{N_{e^+e^-}}{S_\perp\cdot T}=\frac{eE_L\cdot e^2{\cal A}_0^2}{4\pi^2\sqrt{3}}\int_0^1\,d\lambda\,\lambda(1-\lambda)\frac{\frac{1}{2}e^2{\cal A}_0^2(1-2\lambda)^2+8\left[m^2+e^2{\cal A}_0^2\lambda(1-\lambda)\right]}{\left[m^2+e^2{\cal A}_0^2\lambda(1-\lambda)\right]^{3/2}}.
\end{equation}
The residual integration over $\lambda$ is carried out easily with the final result
\begin{equation}
\label{numpairsdelta}
\frac{N_{e^+e^-}}{S_\perp\cdot T}=\frac{m^3 (eE_L/m^2)}{16\pi^2\sqrt{3}}F(\xi_{s}),\quad F(\xi_{s})=2+\left(5\xi_{s}-\frac{4}{\xi_{s}}\right)\,\arctan\left(\frac{\xi_{s}}2\right),\quad \xi_{s}=\frac{e}{m}{\cal A}_0.
\end{equation}
Let us stress that this expression for the pair production rate is exact in the framework of the external background field approach. In particular, no assumptions on the value of the parameter $\xi_{s}$ were presumed. In particular, in the limiting cases we have $F(\xi_{s})\approx (8/3)\xi_{s}^2$ ($\xi_{s}\ll 1$) and $F(\xi_{s})\approx (5\pi/2) \xi_{s}$ ($\xi_{s}\gg 1$), so that
\begin{equation}
\label{delta-limits}
\frac{N_{e^+e^-}}{S_\perp\cdot T}\approx\left\{\begin{array}{ll} \displaystyle \frac{m^3 (eE_L/m^2)}{6\pi^2\sqrt{3}}\xi_{s}^2,& \xi_{s}\ll 1,\\
\displaystyle \frac{5m^3 (eE_L/m^2)}{32\pi\sqrt{3}}\xi_{s},& \xi_{s}\gg 1.
\end{array}\right.
\end{equation}

\begin{figure}
\includegraphics[width=0.49\textwidth]{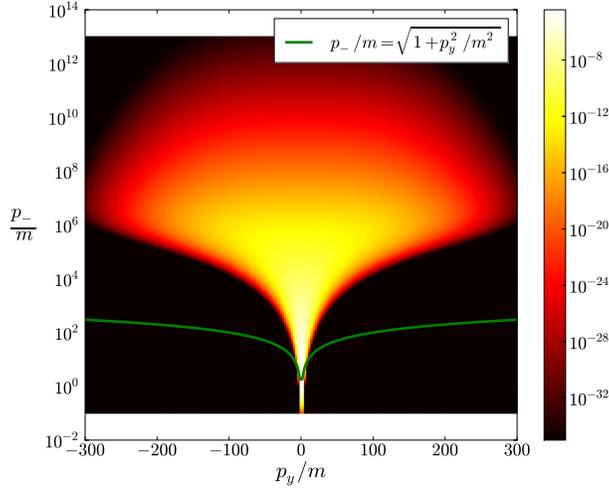}
\caption{Distribution of the number of created particles in quantum numbers $p_-$ and $p_y$, the solid green line separates the upper region $p_z<0$ from the lower region $p_z>0$; field parameters are $\xi_s=1.0$, $E_L/E_c=1.0$.}
\label{distribution:a}
\end{figure}    
		
\begin{figure}
\includegraphics[width=0.49\textwidth]{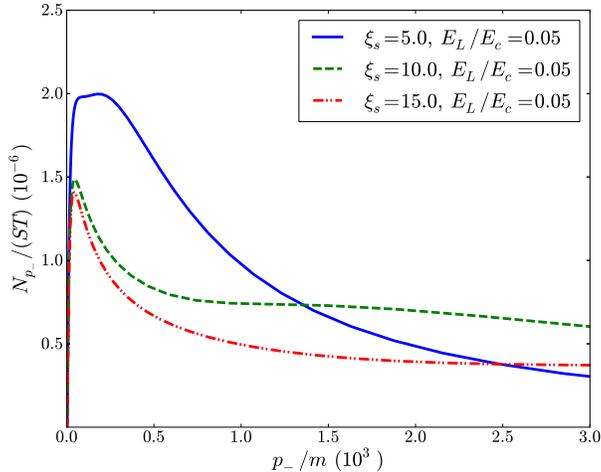}
\caption{Particle production rate per unit wave front area vs. $p_-$ for $\xi_s=5.0;\,10.0;\, 15.0$.}
\label{distribution:b}		
\end{figure}

We could rather integrate in the expression (\ref{numpairs1}) over $q_-$ exclusively and leave it non-integrated over $p_-$ and $p_y$, this would give the momentum distribution of the created particles (both quantities $p_-$ and $p_y$ are conserved in the constant crossed field $A_L$ alone). A typical distribution over $p_-$ and $p_y$ is shown in Fig.~\ref{distribution:a}. This distribution is plotted under the assumption $p_x=0$. One can express $p_z$ in terms of $p_-$ and $p_\perp$: $p_z=\frac{1}{2}\left(-p_-+\frac{m^2+\vec{p}_\perp^2}{p_-}\right)$. According to this expression, particles with $p_->\sqrt{m^2+\vec{p}_\perp^2}$ are travelling against $z$-axis, i.e., are initially carried along with the "hard" pulse. However, as $p_x$ is not conserved and is growing in magnitude, such particles are after all turned back by the field $A_L$. Distribution over $p_-$ only is shown in Fig. \ref{distribution:b} for different values of $\xi_s=e{\cal A}_0/m$. As ${\cal A}_0$ grows, the peak lowers and shifts to the left, but the whole distribution becomes broader.

If one integrates in (\ref{numpairs2}) over $\lambda$ only, then one gets the distribution over $p_y$,
\begin{widetext}
\begin{equation}
\label{numpairs3}
\frac{N_{e^+e^-}}{S_\perp\cdot T}=\frac{eE_L}{4\pi^3\sqrt{3}}\int_{-\infty}^{+\infty} dp_y\; \left[1-
\frac{2(m^2+p_y^2-e^2{\cal A}_0^2/2)}{e{\cal A}_0\sqrt{m^2+p_y^2+e^2{\cal A}_0^2/4}}\,{\rm arth}\,\left({\frac{e{\cal A}_0/2}{\sqrt{m^2+p_y^2+e^2{\cal A}_0^2/4}}}\right)\right].
\end{equation}
\end{widetext}
The shape of distribution over $p_y$ is clear from Fig. \ref{distribution:a}. It is bell-shaped peaking at $p_y=0$ and also gets wider as ${\cal A}_0$ is growing. 

\section{Limiting cases}
\label{sec-comparison}

As was discussed is the previous sections, the two limiting cases for $\xi_{s}$ correspond to the well studied approximations - the perturbative approximation with respect to the ``hard'' pulse, and the locally constant and homogeneous field approximation, so that the expressions (\ref{delta-limits}) can be compared to them. To do that we need first to calculate the number of pairs in the same problem by using these approximations.

Let us consider the case $\xi_{s}\ll 1$, for which the pair creation rate is proportional to the probability of pair creation by a single hard photon in external field $\vec{A}_L$ (perturbative approximation). According to \cite{Ritus}, the probability of scalar pair creation by a single photon with the energy $\omega$ in a constant crossed electromagnetic field $A_L$ per unit time and volume is 
\begin{equation}
\label{Wpar}
W_\parallel(\omega)=-\frac{e^2m^2}{2\omega\sqrt{\varkappa}}\int_{(4/\varkappa)^{2/3}}^\infty d\zeta\,\frac{(\kappa \zeta^{3/2}+2) \,{\rm Ai}'\,(\zeta)}{\zeta^{11/4}\,\sqrt{\varkappa \zeta^{3/2}-4}},\;
\end{equation}
where $\varkappa=eE_Lk_-/m^3=2eE_L\omega/m^3$, and the photon is assumed to be linearly polarized along $A_L$. Since the photon beam is composed of photons with different frequencies, the total number of pairs is given by the integral of (\ref{Wpar}) over the photon spectrum
\begin{equation}
\label{Npar}
\frac{N_{e^+e^-}}{S_\perp\cdot T}=\frac1{S_\perp}\int_0^\infty W_\parallel(\omega)\frac{dN_\gamma(\omega)}{d\omega}\,d\omega .
\end{equation}
Note that this formula obviously corresponds to the first order of perturbation theory with respect to the $\gamma$-pulse. In the case of our interest, the photon beam is described by a delta function, $\vec{E}_\gamma=\vec{{\cal A}}_0\,\delta(x_+)$. One can equate the classical field to the general expression for the operator of electromagnetic field in QED:
\begin{equation}\label{eq}
\vec{E}_\gamma=\vec{{\cal A}}_0\,\delta(x_+)=\sum\limits_\lambda\int\frac{d^3k}{(2\pi)^{3/2}}\frac{(-i\omega_k)\sqrt{4\pi}}{\sqrt{2\omega_k}}\left(\vec{e}_{\vec{k}\lambda}c_{\vec{k}\lambda}e^{-ikx}-\vec{e}_{\vec{k}\lambda}^*c_{\vec{k}\lambda}^\dagger e^{+ikx}\right),
\end{equation}
where $k$ is the photon momentum, $\vec{e}_{\vec{k}\lambda}$ is the polarization vector and $c_{\vec{k}\lambda}$ are the weights of the modes (in QED, photon annihilation operators). From Eq.~(\ref{eq}), we can derive $c_{\vec{k}\lambda}$ required for the chosen field $\vec{E}_\gamma$
\begin{equation}
c_{\vec{k}\lambda}=i\delta_{\lambda,\parallel}\frac{\delta(\vec{k}_\perp)\theta(-k_z)}{\sqrt{\omega_k}}.
\end{equation}
Accordingly, the photon spectrum is of the form
\begin{equation}
\label{spectrum}
dN_\gamma(\omega)=|c_{\vec{k},\parallel}|^2\,d^3k=\frac{S_\perp{\cal A}_0^2\,d\omega}{(2\pi)^2\omega}.
\end{equation}
By substituting (\ref{spectrum}) into (\ref{Npar}) and passing to the new integration variable $\omega\,\rightarrow\,\varkappa=2eE_L\omega/m^3$, we obtain
\begin{equation}
\frac{N_{e^+e^-}}{S_\perp\cdot T}=-\frac{(eE_L)(e{\cal A}_0)^2}{(2\pi)^2m}\int_0^\infty\frac{d\varkappa}{\varkappa^{5/2}}\int_{(4/\varkappa)^{2/3}}^\infty d\zeta\,\frac{(\varkappa \zeta^{3/2}+2)\,{\rm Ai}'\,(\zeta)}{\zeta^{11/4}\,\sqrt{\varkappa \zeta^{3/2}-4}}.
\end{equation}
After swapping the order of integrations, the integral over $\varkappa$ is computed easily, 
\begin{equation}
\frac{N_{e^+e^-}}{S_\perp\cdot T}=-\frac{(eE_L)(e{\cal A}_0)^2}{(2\pi)^2m}\int_0^\infty \frac{d\zeta}{\zeta^{11/4}}\,{\rm Ai}'\,(\zeta)\frac{2}{3}\zeta^{9/4}
\end{equation}
Using the formula $\int_0^\infty \frac{d\zeta}{\sqrt{\zeta}}\,{\rm Ai}'\,(\zeta)=-\frac1{\sqrt{3}}$, we finally obtain the total number of pairs
\begin{equation}
\frac{N_{e^+e^-}}{S_\perp\cdot T}=\frac{(eE_L)(e{\cal A}_0)^2}{6\pi^2\sqrt{3}\,m},\quad \xi_{s}\ll 1,
\end{equation}
which exactly coincides with the corresponding limiting case in (\ref{delta-limits}). In other words, in the limit $\xi_{s}\ll 1$ the exact solution (\ref{numpairsdelta}) describes pair creation by a dilute photon gas in external constant crossed field, according to the first order of perturbation theory with respect to $\gamma$-pulse. The number of pairs created is proportional to the average number of photons in the "hard" pulse $N_{e^+e^-}\propto \xi_{s}^2\propto \bar N_\gamma$, which is rather understandable.

Now let us pass to the limiting case $\xi_{s}=e{\cal A}_0/m=eE_\gamma\tau/m\gg 1$. In this case, let us try the locally constant field approximation, which means that the resulting field of superposition of both pulses is assumed to be nearly constant at the neighbourhood of each spatial point. In this approximation, the number of scalar pairs can be computed as follows \cite{Ritus,focused}
\begin{equation}
N_{e^+e^-}=\int dt\,dV \frac{m^4\varepsilon(z,t)\eta(z,t)}{8\pi^2\sinh\left[\pi\eta(z,t)/\varepsilon(z,t)\right]}
\exp\left(-\frac{\pi}{\varepsilon(z,t)}\right).
\end{equation}
In our case, the invariant parameter $\eta=0$, so that
\begin{equation}
N_{e^+e^-}=\int dt\,dV \frac{m^4\varepsilon^2(z,t)}{8\pi^3}\,\exp\left(-\frac{\pi}{\varepsilon(z,t)}\right).
\end{equation}
If we substitute $\varepsilon=2e\sqrt{E_L {\cal A}_0\delta(x_+)}/m^2$ and take into account that $\exp \left(-\pi/\varepsilon(x_+)\right)\vert_{x_+=0}=1$ and $\exp \left(-\pi/\varepsilon(x_+)\right)\vert_{x_+\neq 0}=0$, then we have
\begin{equation}
N_{e^+e^-}=\int dt\,dV \frac{e^2E_L{\cal A}_0}{2\pi^3}\delta(x_+)\sim \frac{1}{2\pi^3}(eE_L)(e{\cal A}_0)S_\perp T.
\end{equation}
Note that surprisingly the functional dependence on the parameters of the field here is the same as for the limiting case $\xi_{s}\gg 1$ in (\ref{delta-limits}). However, there remains some difference in a numerical coefficient: in the locally constant field approximation it is $(2\pi^3)^{-1}\approx 1.61\times 10^{-2}$ instead of $5/(32\pi\sqrt{3})\approx 2.87\times 10^{-2}$ in (\ref{delta-limits}). Of course, this should not be regarded a problem, because the locally constant field approximation can not be supposed to be accurate for our model with a delta-pulse.

\section{Discussion}
\label{sec-conclusion}

The problem of pair creation by external electromagnetic field of two counterpropagating plane wave pulses is of great interest for planning or discussing experiments but generally does not admit exact analytical solutions. If both pulses are optical and strong ($\xi_L,\,\xi_s\gg 1$), then quantitative estimates of pair creation can be satisfactorily obtained by applying the locally constant field approximation \cite{focused, colliding1, colliding2}. But an even more interesting setup would be a collision of an ultra-strong optical laser pulse with an ultra-short pulse. The latter is composed of high-frequency photons and, as is widely expected \cite{Dunne}, pair creation could benefit from it. If $\xi_s\ll 1$, then the ultra-short pulse can be well considered as a dilute gas of individual mutually incoherent hard photons, and the problem can be solved perturbatively with respect to its field. However, there is a problem with the transient region ($\xi_s\sim 1$), where the only currently available approach is the imaginary time method, but the latter is capable for analytical quantitative results only if the spatial variation of the field is completely ignored, which seems to be not a reasonable assumption\footnote{For example, the well known fact that a single plane wave of arbitrary shape does not create pairs obviously can not be explained within such an approach}.

In the present paper we first of all demonstrate how to obtain the already known formulas (\ref{w-electric-L}), (\ref{w-electric}) for pair photoproduction in a constant electric field in a rather simple way by a sort of quasiclassical method. It turns out that the structure of the exponential factor in the probability of pair photoproduction can be understood almost as simply as that of the exponential factor in the probability of vacuum pair production. These results can be readily generalized to the case of pair production by a photon in an arbitrary constant electromagnetic field (though, for the sake of simplicity, we restrict ourselves to a simpler case $\eta=0$). We also discuss a conjecture on how to extend the area of applicability of these formulas to a more realistic setup with two counterpropagating pulses, at least without aiming to compute a preexponential factor, by proceeding in a spirit of the mean-field approach. Namely, it may occur reasonable to take into account the variation of the field of the ``hard'' pulse by picking out a single representative hard photon from it and consideration of probability of pair photoproduction of this photon in a total (``mean'') field of superposition of the two pulses, which is assumed constant. Intuitively, this would correspond to a first term in a virial expansion of the probability in powers of $\xi_s$. The results of this part confirm the widely discussed expectations that usage of ``hard'' pulses (i.e., combining the high-power with high frequency lasers) may enlarge the pair production yield.

Besides, we introduce a new type of exactly solvable models for pair creation, with one of the colliding pulses been represented by a delta-pulse. In this case, pairs are created exclusively inside the overlap region (i.e., inside the delta-pulse), so that it becomes possible to identify the in- and out-regions and to obtain explicitly the Bogolubov transformation connecting the in- and out- creation and annihilation operators of a charged field. In this particular case, the Bogolubov transformation arises as a matching condition at a surface of the wave front of the delta-pulse. Of course, the models with a delta-pulse can be hardly applied directly to estimates relevant for realistic experimental setup, because they correspond to an infinite (supercritical) field strength inside a ``hard'' pulse and, as a result, the most important factor of the probability -- the tunneling exponential suppression factor -- is lacking at all. However, the parameter $\xi_s$ can be redefined for this case and dependence on it (unlike the other known approaches) can be traced exactly. We hope that this goal (as well as the whole structure of the Bogolubov transformation, which turns out to be rather non-trivial) will give some further clues on the nature of non-perturbative regime of pair production. In the current paper, we restrict ourselves just to discussion of the main features of the resulting expressions. 

One of the features of the result is that the pair production rate is proportional to the mean number of photons $\bar N_\gamma$ in a delta-pulse (equivalently, intensity, or photon density) only in the limit $\xi_s\ll 1$, which can be also considered perturbatively with respect to the field of a delta-pulse. The opposite limiting case $\xi_s\gg 1$, in a sense, corresponds to the locally constant field approximation, although it can be hardly supposed to be valid literally for non-continues fields. In this regime, $N_{e^+e^-}\propto \xi_{s}\propto \sqrt{\bar N_\gamma}$, which may resemble a spread of the Poisson distribution (as is well known, the classical external field corresponds to coherent states of the quantum field, which possess Poissonian statistics). Since this regime is non-perturbative with respect to a delta-pulse, a possible explanation is as follows: during a pair creation process some of the ``hard'' photons are absorbed and some are emitted, but harder photons are mostly absorbed while softer photons are mostly emitted, so that the net number of absorbed photons is proportional to the spread of the photon energy distribution. But maybe this result should be explained more accurately.

In this paper we elucidated pair creation in a model with a delta-pulse in a simplest situation -- we considered creation of scalar pairs, in a collision of a constant crossed field with a single linearly polarized delta-pulse. However, the model maintains exact solubility with many generalizations, e.g. for fermion instead of scalar pair production, with a replacement of a single delta-pulse by trains of arbitrarily polarized delta-pulses and of a constant crossed field by an arbitrary plane wave field. Generalization to the case of inclined collision seems to be also possible. We believe that further development of this approach will be useful for understanding of some aspects of non-perturbative regime of QFT.

\begin{acknowledgments}
The work was supported by the Russian Fund for Basic Research (grants 11-02-12148ofi-m and 13-02-00372), the Ministry of Science and Education of the Russian Federation within the Federal Program ``Scientific and scientific-pedagogical personnel of innovative Russia 2009-2013'' (agreement 14.A18.21.0773), and the President program for support of young Russian scientists and leading research schools (grant no. MD-5838.2013.2). We are also grateful to N.B.~Narozhny for valuable discussions.
\end{acknowledgments}

\appendix*
\section{Orthogonality and normalization of Volkov states}
\label{appendix}
For completeness, let us sketch the proof of the orthogonality and normalization condition(\ref{normalization}) for the Volkov solutions (\ref{Volkov}). To do this, it is more convenient to rewrite (\ref{explicit-Volkov}) in the following way:
\begin{equation}
\label{explicit-Volkov2}
\begin{split}
\Psi_{\vec{p}_\perp,\,p_-}=&\frac1{\sqrt{(2\pi)^32|p_-|}}\,\exp\left\{i\vec{p}_\perp\cdot\vec{x}_\perp-i\frac{p_-x_+}{2}\vphantom{\int\limits}\right.\\&\left.-\frac{i}{2p_-}\int_0^{x_-}\left[\left(\vec{p}_\perp+e\vec{A}_L(x_-)\right)^2+m^2\right]\,dx_-\right\}.
\end{split}
\end{equation}
In our gauge $A_0=A_3=0$, so that $A_+=0$. Hence, after substitution of (\ref{explicit-Volkov2}) into (\ref{normalization}), we have
\begin{equation}
\begin{split}
\int d^2x_\perp dx_- \Psi_{\vec{p}_\perp,p_-}^*i\stackrel{\leftrightarrow}{\frac{\partial}{\partial x_-}}\Psi_{\vec{p}^{\,\prime}_\perp,p_-'}=\frac{1}{(2\pi)^3 2\sqrt{|p_-||p_-'|}}\int d^2x_\perp dx_-\,Q(x,\,p)e^{iS(x,\,p)},
\end{split}
\end{equation}
where
\begin{equation}
Q(x,\,p)=\frac{\left({\vec p}_\perp^{\,\prime}+e\vec{A}_L(x_-)\right)^2+m^2}{2p_-'}+\frac{\left({\vec p}_\perp +e\vec{A}_L(x_-)\right)^2+m^2}{2p_-}
\end{equation}
and
\begin{equation}
\begin{split}
S(x,\,p)=&({\vec p}_\perp^{\,\prime}-{\vec p}_\perp)\cdot\vec{x}_\perp-i(p_-'-p_-)\frac{x_+}{2}\\
&-\int_0^{x_-}\left[\frac{\left(\vec{p}_\perp^{\,\prime}+e\vec{A}_L(x_-)\right)^2+m^2}{2p_-'}-\frac{\left(\vec{p}_\perp+e\vec{A}_L(x_-)\right)^2+m^2}{2p_-}\right]\,dx_-.
\end{split}
\end{equation}
Integration over ${\vec x}_\perp$ results in $(2\pi)^2\delta(\vec{p}_\perp^{\,\prime}-\vec{p}_\perp)$, and on account of it we obtain
\begin{equation}\label{app}
\begin{split}
\int d^2x_\perp &dx_- \Psi_{\vec{p}_\perp,p_-}^*i\stackrel{\leftrightarrow}{\frac{\partial}{\partial x_-}}\Psi_{\vec{p}^{\,\prime}_\perp,p_-'}=\frac{\delta(\vec{p}_\perp^{\,\prime}-\vec{p}_\perp)}{2\pi\cdot 2\sqrt{|p_-||p_-'|}}\int dx_- \frac{1}{2}\left[({\vec p}_\perp+e\vec{A}_L(x_-))^2+m^2\right] \\
\times &\frac{p_-+p_-'}{p_-p_-'}\exp\left\{-i(p_-'-p_-)\frac{x_+}{2}-i\frac{p_--p_-'}{2p_-p_-'}\int_0^{x_-}\left[\left(\vec{p}_\perp^{\,\prime}+e\vec{A}_L(x_-)\right)^2+m^2\right]dx_-\right\}.
\end{split}
\end{equation}
Finally, let us change the integration variable $x_-\rightarrow u(x_-)$, where
\begin{equation}
u(x_-)=\frac{1}{2p_-p_-'}\int_0^{x_-}\left[\left(\vec{p}_\perp^{\,\prime}+e\vec{A}_L(x_-)\right)^2+m^2\right]dx_-.
\end{equation}
Then (\ref{app}) takes the form
\begin{equation}
\begin{split}
\int d^2x_\perp dx_- \Psi_{\vec{p}_\perp,p_-}^*i\stackrel{\leftrightarrow}{\frac{\partial}{\partial x_-}}\Psi_{\vec{p}^{\,\prime}_\perp,p_-'}=&\delta(\vec{p}_\perp^{\,\prime}-\vec{p}_\perp)\frac{(p_-+p_-')}{2\pi\cdot 2\sqrt{|p_-||p_-'|}}\\
&\times\int du\,\exp\left[-i(p_-'-p_-)\frac{x_+}{2}-i(p_-'-p_-)u\right].
\end{split}
\end{equation}
By noting that integration over $u$ gives $(2\pi)\delta(p_-'-p_-)$, and taking into account that $\frac{p_-}{|p_-|}={\rm sgn}(p_-)$, we arrive at Eq.~(\ref{normalization}).

\end{document}